\documentclass[
journal=jacsat, 
manuscript=article]{achemso}

\usepackage[version=3]{mhchem} 

\author{Bodo Zibrowius}
\email{bodo.zibrowius@posteo.de}
\affiliation
{45468 M{\"u}lheim an der Ruhr, Germany}
\author{Michael Fischer}
\email{michael.fischer@uni-bremen.de}
\affiliation
{Crystallography \& Geomaterials, Faculty of Geosciences, University of Bremen, Klagenfurter Stra{\ss}e
2-4, 28359 Bremen, Germany}
\alsoaffiliation{Bremen Center for Computational Materials Science, University of Bremen, 28359 Bremen, Germany}
\alsoaffiliation{MAPEX Center for Materials and Processes, University of Bremen, 28359 Bremen, Germany}

\title[\texttt{achemso} demonstration]
{On the use of Solomon echoes in $^{27}$Al NMR studies of complex aluminium hydrides}

\begin{document}

\begin{abstract}
The quadrupole coupling constant $C_Q$ and the asymmetry parameter $\eta$ have been determined for two complex aluminium hydrides from $^{27}$Al NMR spectra recorded for stationary samples by using the Solomon echo sequence. The thus obtained data for \ce{KAlH4} ($C_Q=(1.30\pm0.02)$\,MHz, $\eta=(0.64\pm0.02)$) and \ce{NaAlH4} ($C_Q=(3.11\pm0.02)$\,MHz, $\eta<0.01$) agree very well with data previously determined from MAS NMR spectra. The accuracy with which these parameters can be determined from static spectra turned out to be at least as good as via the MAS approach.
The experimentally determined parameters ($\delta_{iso}$, $C_Q$ and $\eta$) are compared with those obtained from DFT-GIPAW (density functional theory - gauge-including projected augmented wave) calculations. Except for the quadrupole coupling constant for \ce{KAlH4}, which is overestimated in the GIPAW calculations by about 30\%, the agreement is excellent. Advantages of the application of the Solomon echo sequence for the measurement of less stable materials or for \textit{in-situ} studies are discussed. 
\end{abstract}

\section{Introduction}

Complex aluminium hydrides have been studied rather extensively  in the past two decades, mainly because of their potential application as hydrogen storage materials. Triggered by the seminal paper by  Bogdanovi\'{c} and Schwickardi \cite{Bogdanovic97}, the reversible dehydrogenation of \ce{NaAlH4} under the influence of catalysts was the target of many experimental and theoretical studies. In the course of these investigations, a large number of materials containing aluminium hydrides have been proposed for the reversible storage of hydrogen. Several new complex aluminium hydrides have been discovered and characterized. The progress in this research area has regularly been reviewed from different perspectives.\cite{Schueth04, Orimo07, Bogdanovic09, Frankcombe12, Li13, Callini16, Milanese18, Suarez19, Zhao21}

Beside diffraction methods, NMR spectroscopy has been proven to be a valuable tool for the identification and quantification of the various aluminium hydrides that might be present in the samples under study.\cite{Kellberg90, Tarasov97, Tarasov00, Bogdanovic03, Wiench04, Hwang07, Kabbour07, Zhang09, Verkuijlen09, Verkuijlen10, Verkuijlen11, Krech14, Nielsen14, Ares16, ZF19} Most of the $^{27}$Al NMR spectra reported for aluminium hydrides were recorded under MAS conditions, i.e.\ for samples spinning fast around an axis that is inclined by an angle of 54$^\circ$\,44\char"27~ relative to the direction of the magnetic field. This technique is one of the standard methods to tackle the resolution problems in solid-state NMR spectroscopy. It significantly reduces the line  broadening caused by second-order quadrupole interaction and removes the broadening brought about by the heteronuclear dipole-dipole interaction as well as by the anisotropy of the chemical shift. \cite{FreudeHaase} The thus enhanced spectral resolution allows the aluminium hydride species to be identified by the position of the centreband of the central transition. It should be noted that there are cases were this approach is bound to fail. If the quadrupole interaction is as strong as for instance recently found for the aluminium nuclei in \ce{Mg(AlH4)2}\cite{Mamatha206} or in an alane amine adduct, \cite{Ortmeyer19} the static linewidth of the central transition at the usually available magnetic fields exceeds the available spinning speeds by far. Nevertheless, $^{27}$Al NMR spectra can be recorded for non-spinning samples by an echo technique and can be used to characterize these materials. 

Quadrupole interaction should not so much be regarded as an obstacle for recording highly resolved NMR spectra of solids, but first and foremost as a valuable source of information difficult to be gathered by other means. Any nucleus with a spin $I> \tfrac{1}{2}$ has a non-vanishing electric quadrupole moment that interacts with the electric field gradient (efg) at the site of the nucleus. This interaction influences the NMR frequency of the nucleus studied. Hence, any quadrupole nucleus can be regarded as a probe of the local geometry in general and of the symmetry in particular. Apart from the orientation of its principal axes system, the quadrupole coupling can be fully described by just two parameters: the so-called quadrupole coupling constant $C_Q$, which is proportional to both the strength of the gradient and the electric quadrupole moment of the nucleus under study, and the asymmetry parameter $\eta$ of the efg.\cite{FreudeHaase, quadnmr, Man11} For example, if the nucleus is located on a symmetry axis $C_n$ with $n\geq3$, the efg is axially symmetric and $\eta$ is zero. In a cubic environment, there is no gradient and hence there is no quadrupole interaction.

For materials where the quadrupole interaction is small (or medium sized), the second-order quadrupolar broadening of the central transition is hardly noticeable or causes only a small splitting of a few ppm in the $^{27}$Al MAS NMR spectra. However, these spectra contain much more information than only the isotropic chemical shift. The range over which the spinning sidebands of the satellite transitions spread and the characteristic intensity modulations of these sidebands allow the parameters of the quadrupole interaction to be determined very precisely. This has been shown for \ce{NaAlH4} \cite{Zhang09} and \ce{KAlH4}. \cite{ZF19} 

The major aim of the present paper is to demonstrate that the information about the chemical shift and the quadrupole interaction can alternatively be gained from $^{27}$Al NMR spectra obtained from Solomon echoes recorded for non-spinning samples. Although the formation of these echoes was demonstrated for the first time more than 60 years ago \cite{Solomon58} and their origin is well understood,\cite{Man97, Man00} Solomon echoes have rarely been used to observe the satellite transitions in solid-state  $^{27}$Al NMR spectroscopy.\cite{Azais02} The precision of NMR data obtained from stationary samples is generally assumed to be lower than of those derived from MAS NMR spectra. However, this approach avoiding fast sample spinning is certainly advisable for less stable materials that are prone to  decomposition under mechanical and/or thermal stress.

With the above discussed high sensitivity of the quadrupole interaction to the local environment of the nucleus, the precise determination of the quadrupole coupling parameters combined with DFT (density functional theory) calculations can offer an alternative approach to high-quality structures for polycrystalline materials. \cite{Perras12} We show that the experimental data obtained for \ce{NaAlH4} and \ce{KAlH4} are well reproduced by DFT-GIPAW (gauge-including projected augmented wave) calculations.\cite{Bonhomme2012}

\section{Experimental and computational methods}

\subsection{Materials}

\ce{NaAlH4} (Chemetall, 82–-85\%) was purified by dissolving it in tetrahydrofurane and filtrating off the insoluble portions. The pure \ce{NaAlH4} was precipitated from the solution by addition of pentane and carefully dried in vacuum.

 \ce{KAlH4} was produced by ball milling NaAlH$_{4}$ and KCl.\cite{ZF19} The resulting powder was suspended in diglyme. After filtration, KAlH$_{4}$ was precipitated through the addition of toluene and filtered off. The dried KAlH$_{4}$ contained small amounts of unreacted NaAlH$_{4}$ which were removed by treatment in tetrahydrofurane.

	All syntheses and operations were performed under argon using dried and oxygen-free solvents. The MAS rotors were filled and capped in a glove box and transferred to the spectrometer in argon-filled vials.

\subsection{Solid-state NMR spectroscopy}

The $^{27}$Al NMR spectra were recorded on a Bruker Avance\,III\,HD 500WB spectrometer using double-bearing MAS probes (DVT BL4) at resonance frequencies of 130.3\,MHz. The chemical shift was referenced relative to an external 1.0\,M aqueous solution of aluminium nitrate. The same solution was used for determining the flip-angle. 

For the $^{27}$Al MAS~NMR spectra, single $\pi$/12 pulses ($t_p$ = 0.6\,$\mu$s) were applied at a repetition time of 2\,s (2,000--16,000 scans) and spinning frequencies ($\nu_{MAS}$) between 3 and 13\,kHz.  High-power proton decoupling (SPINAL-64) was used for all $^{27}$Al~NMR spectra shown in this paper. The magic angle was adjusted by maximizing the rotational echoes of the $^{23}$Na resonance of solid \ce{NaNO3}.

The $^{27}$Al NMR spectra of stationary samples were acquired using the Solomon echo sequence with two pulses of the same length $t_p$ separated by a delay $\tau$.\cite{Solomon58} For all half-integer spins with $I > \tfrac {3}{2}$, this pulse sequence generates in general a whole series of echoes. For $^{27}$Al ($I = \tfrac {5}{2}$), echoes at $\tau$, 2$\tau$, and 3$\tau$ after the second pulse are expected for the inner satellite transitions ($(+\tfrac{1}{2}\leftrightarrow+\tfrac{3}{2})$ and $(-\tfrac{3}{2}\leftrightarrow-\tfrac{1}{2})$) and at $\tau$/2, $\tau$, and 3$\tau$/2 for the outer satellite transitions ($(+\tfrac{3}{2}\leftrightarrow+\tfrac{5}{2})$ and  $(-\tfrac{5}{2}\leftrightarrow-\tfrac{3}{2})$).\cite{Man97}  The echoes at  $\tau$/2, $\tau$, and 2$\tau$ are referred to as allowed echoes and those at 3$\tau$ and 3$\tau$/2 are referred to as forbidden echoes.\cite{Solomon58, Man97, Man00} Which of the various echoes are experimentally observed can to a certain extent be influenced by the phase cycling applied.\cite{Bonhomme04} The echo at $\tau$ that is used here to obtain the spectra contains the spectral information of both satellite transitions. Solomon echoes can be generated with two in-phase pulses ($p_{x}$--$\tau$--$p_{x}$--$\tau$--acq) or two pulses in quadrature phase ($p_{x}$--$\tau$--$p_{y}$--$\tau$--acq).\cite{Man92, Man97} Combining both variants with phase alternation of the second pulse and CYCLOPS (CYCLically Ordered Phase Sequence\cite{Hoult75}) yields the 16-step phase cycle that was originally proposed by Kunwar et al. \cite{Kunwar86} This phase cycle, which is known to effectively cancel spurious signals from the NMR probe,\cite{Man92} was used for the acquisition of all Solomon echo data shown here.

\begin{figure}
  \includegraphics[width=8.3cm]{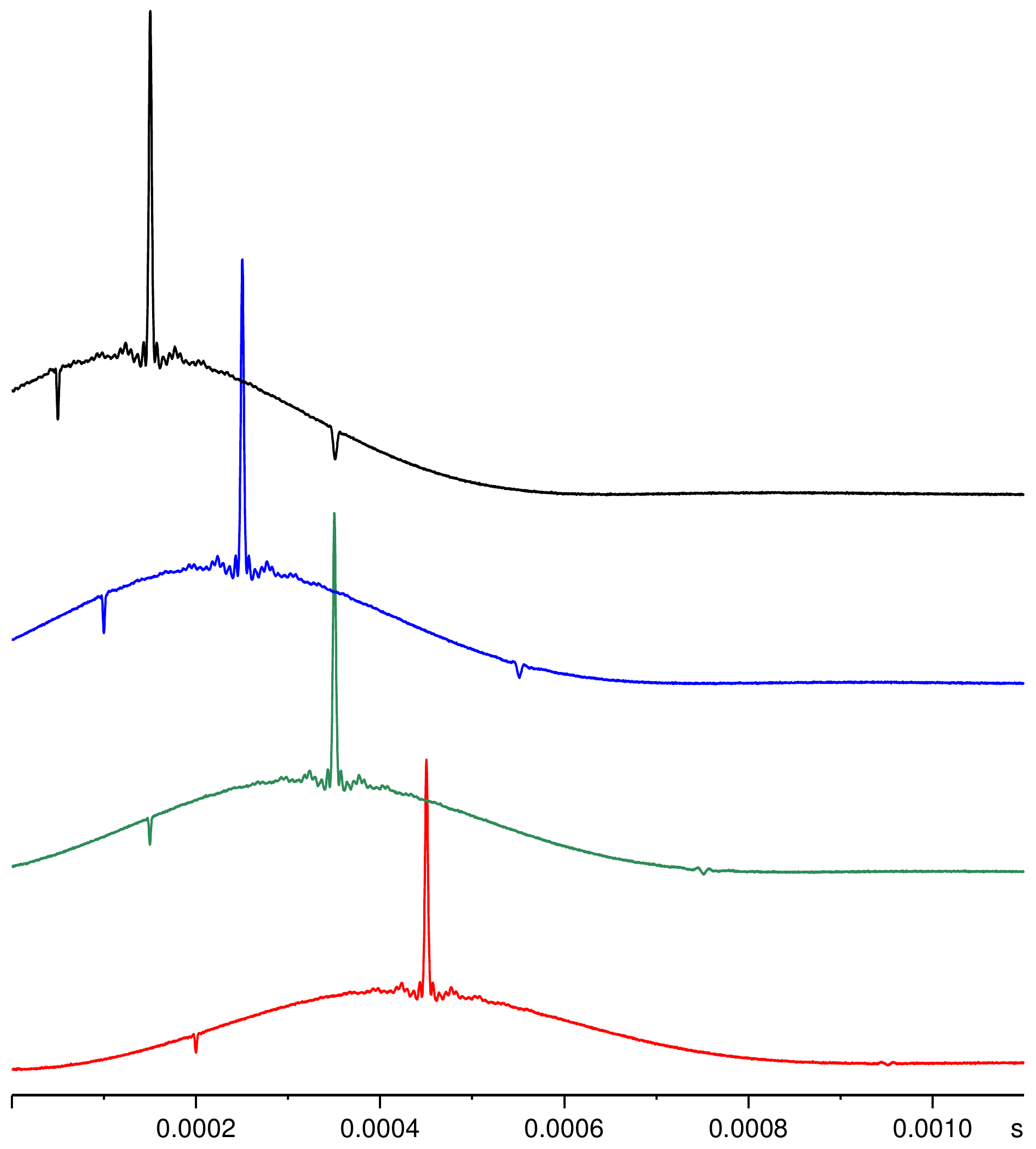}
  \caption{$^{27}$Al NMR Solomon echoes generated for \ce{KAlH4} by applying two pulses ($t_p$ = 0.9\,$\mu$s) separated by 200, 300, 400,  and 500\,$\mu$s (from top to bottom). For recording these echoes, a short pre-acquisition delay of only 50\,$\mu$s was used.} 
  \label{fgr:Echoes}
\end{figure}

Solomon echoes that were obtained in this way for \ce{KAlH4}  with various pulse spacings $\tau$  are shown in Fig.~\ref{fgr:Echoes}. The additional echoes at  $\tau$/2 and 2$\tau$ can easily been discerned. The echo at $\tau$/2 is narrower than the other two since it contains only the spectral information from the wider outer satellite transition. Obviously, the forbidden echo at 3$\tau$/2 is not observed under the experimental conditions chosen. This also holds true for the forbidden echo at 3$\tau$.
The existence of several echoes for nuclei with $I \geq \tfrac {5}{2}$ is generally regarded as a main drawback of this technique since Fourier transform of the time signal generally leads to distorted spectra,  necessitating a direct analysis of the echoes in the time domain. \cite{Man00}

To minimize the interfering effect of the echo generated at 2$\tau$ after the second pulse, we generally used pulse spacings $\tau$ of at least 0.5\,ms. For both \ce{KAlH4} and \ce{NaAlH4}, intense echoes at $\tau$ could be obtained for pulse spacings up to 1.2\,ms. Since Solomon echoes can only 
be observed if the pulse spacing is much smaller than the duration of the free induction decay of the central transition ($T_{FID}$),\cite{Man97, Man00} the rather slow spin-spin relaxation in these two aluminium hydrides is obviously a fortunate circumstance for the application of this technique.  

For the spectra shown here, two strong rf pulses ($\nu_{rf}$ $\approx$ 100\,kHz) with a length $t_p$ = 0.9\,$\mu$s were applied. With a repetition time of 2\,s, between 16,000 and 48,000 scans were accumulated. To start Fourier transform at the top of the echo at $\tau$, a pre-acquisition delay slightly shorter than $\tau$ and an appropriate number of left shifts were applied (dwell time: 0.05\,$\mu$s). 

The spectra simulations were performed using the solids lineshape analysis module implemented in the TopSpin\texttrademark\ 3.2 NMR software package from Bruker BioSpin GmbH.

\subsection{DFT-GIPAW calculations}

All DFT calculations were carried out using the CASTEP code.\cite{Clark2005,Milman2010} The calculations employed the PBE exchange-correlation functional \cite{Perdew1996}, using ultrasoft pseudopotentials generated on the fly and a plane wave cutoff energy of 800 eV. The structures of \ce{NaAlH4} and \ce{KAlH4} were optimized, relaxing all atomic positions while fixing the unit cell parameters to experimental values. The optimizations, which employed a BFGS optimizer, were considered converged when the maximal residual force on an atom fell below 0.001\;eV/\AA\;and when the maximal atomic displacement with respect to the previous step was smaller than 0.0005\;\AA.

The starting structure of \ce{NaAlH4} was taken from the work of Hauback et al. (space group $I4_1/a$, $a$ = 5.0119\;\AA, $c$ = 11.3147\;\AA).\cite{Hauback03} The alternative use of the structural data published by Ozolins et al.\ (space group $I4_1/a$, $a$ = 5.0099\;\AA, $c$ = 11.3228\;\AA)\cite{Ozolins2004} led to essentially identical results. A $7\times7\times3$ mesh of $k$-points, corresponding to 26 irreducible points, was used to sample the first Brillouin zone. For \ce{KAlH4}, the atomic coordinates were taken from a previous neutron diffraction study of \ce{KAlD4},\cite{Hauback05} whereas cell parameters were taken from the more recent X-ray diffraction study of \ce{KAlH4} by Zibrowius and Felderhoff (space group $Pnma$, $a$ = 8.8475\;\AA, $b$ = 5.8143\;\AA, $c$ = 7.3448\;\AA).\cite{ZF19}  For this structure, a $4 \times 6 \times 5$ mesh of $k$-points, corresponding to 18 irreducible points, was used.

DFT calculations of the isotropic shielding parameter $\sigma_{iso}$,  the quadrupole coupling constant $C_Q$, and the asymmetry parameter $\eta$ employed the gauge-including projector augmented wave (GIPAW) method as implemented in CASTEP.\cite{Pickard2001,Profeta2003,Yates2007,Bonhomme2012} Calculations were performed on the DFT-optimized structures, as a prior optimization of the atomic coordinates obtained from diffraction data is crucial to obtain meaningful results for hydrogen-containing systems. Even when neutron diffraction is used, a highly accurate determination of the hydrogen positions is challenging. Inaccuracies in  the hydrogen positions result in large errors in the calculated NMR parameters. \cite{Yates2005,Bonhomme2012,Ashbrook2016} In terms of exchange-correlation functional, pseudopotentials, cutoff energy, and $k$-meshes, the same settings as for the optimisation were used for the DFT-GIPAW calculations. The analysis of the calculated NMR parameters made use of the MagResView tool.\cite{Sturniolo2016}

In order to compare the isotropic magnetic shielding $\sigma_{iso,DFT}$ directly obtained from the DFT-GIPAW computations with the experimentally accessible  chemical shift data, a conversion to $\delta_{iso,DFT}$ is required. In the first place, such a conversion was made by using \ce{LiAlH4} as the sole reference system. Calculations analogous to those described above were carried out for \ce{LiAlH4}, using the crystal structure of \ce{LiAlD4} reported by Hauback et al. \cite{Hauback02} as starting point ($k$-mesh: $8\times5\times5$). The chemical shift of the systems of interest was then calculated as: 
\begin{equation}
  \delta_{iso,DFT,1}=454.8\,\text{ppm} + 102.0\,\text{ppm} - \sigma_{iso,DFT}=556.8\,\text{ppm} - \sigma_{iso,DFT}\text{,}
   \label{eq:delta1}
\end{equation}
where 454.8\,ppm corresponds to the shielding value $\sigma_{iso,DFT}$ calculated for \ce{LiAlH4} and 102.0\,ppm is the isotropic shift experimentally observed with respect to the usual standard.\cite{Kellberg90}

To evaluate the performance of the DFT-GIPAW computations across a broader set of systems, additional calculations were carried out for three other alkali aluminium hydrides, namely \ce{Li3AlH6} ($k$-mesh: $6\times6\times4$),\cite{Brinks2003} \ce{Na3AlH6} ($k$-mesh: $6\times6\times4$),\cite{Ozolins2004} and \ce{Na2LiAlH6} ($k$-mesh: $6\times6\times6$).\cite{Brinks2005}. The references cited for the structures are combined X-ray and neutron diffraction studies of the corresponding isostructural aluminium deuterides. The experimental data for $\delta_{iso}$ of \ce{Li3AlH6} were taken from Wiench et al.\cite{Wiench04} and those for \ce{Na3AlH6} and \ce{Na2LiAlH6} from Zhang et al.\cite{Zhang09}

It should be noted that there are two non-equivalent Al atoms in the rhombohedral crystal structure reported for \ce{Li3AlH6} by  Brinks and Hauback. \cite{Brinks2003} In particular, because of the results of a detailed analysis carried out by L{\o}vvik et al.\cite{Lovvik04,Lovvik05}, we regard this structural solution as rather convincing. Surprisingly, only a single resonance line was observed in the NMR spectrum of \ce{Li3AlH6}.\cite{Wiench04} Our calculations yield a clue to explain this apparent discrepancy. For the magnetic shielding $\sigma_{iso,DFT}$ of Al1 and Al2 in \ce{Li3AlH6}, we obtained 600.2\,ppm and 602.7\,ppm, respectively. The results for the quadrupole coupling constant $C_Q$ are 2.54\,MHz and $-1.54$\,MHz for Al1 and Al2, respectively. For symmetry reasons, the asymmetry parameter is $\eta=0$ for both positions. We assume that the experimentally determined values $\delta_{iso}= -33.7$\,ppm and $C_Q=1.4$\,MHz can be assigned to Al2. According to our calculations, Al1 should give rise to a resonance line about 2.5\,ppm downfield of the line of Al2. However, the much stronger quadrupole coupling leads to a more significant second-order broadening and to a more pronounced quadrupole-induced highfield shift. At the resonance frequency of 104\,MHz used by Wiench et al.\cite{Wiench04}, the centres of gravity of both resonance lines are expected to be only 0.2\,ppm apart of each other. Furthermore, it has to be taken into account that the \ce{Li3AlH6} studied by these authors was produced by ball-milling \ce{LiAlH4} and \ce{LiH}. Immediate products of  mechanochemical syntheses very often have a rather poor crystallinity. 
This seems to be the case for \ce{Li3AlH6} produced via ball milling as the  results of a study devoted to this solid-state phase transformation show.\cite{Balema00} The resolution of the X-ray powder pattern for the obtained product was insufficient to determine the space group. Fig.~S1 in the Supporting Information demonstrates that under these circumstances the resonance line of Al1 would be hardly visible.  A high-field NMR investigation using well crystallized material should be able to verify our hypothesis. 

\begin{figure}
	\includegraphics[width=8.3cm]{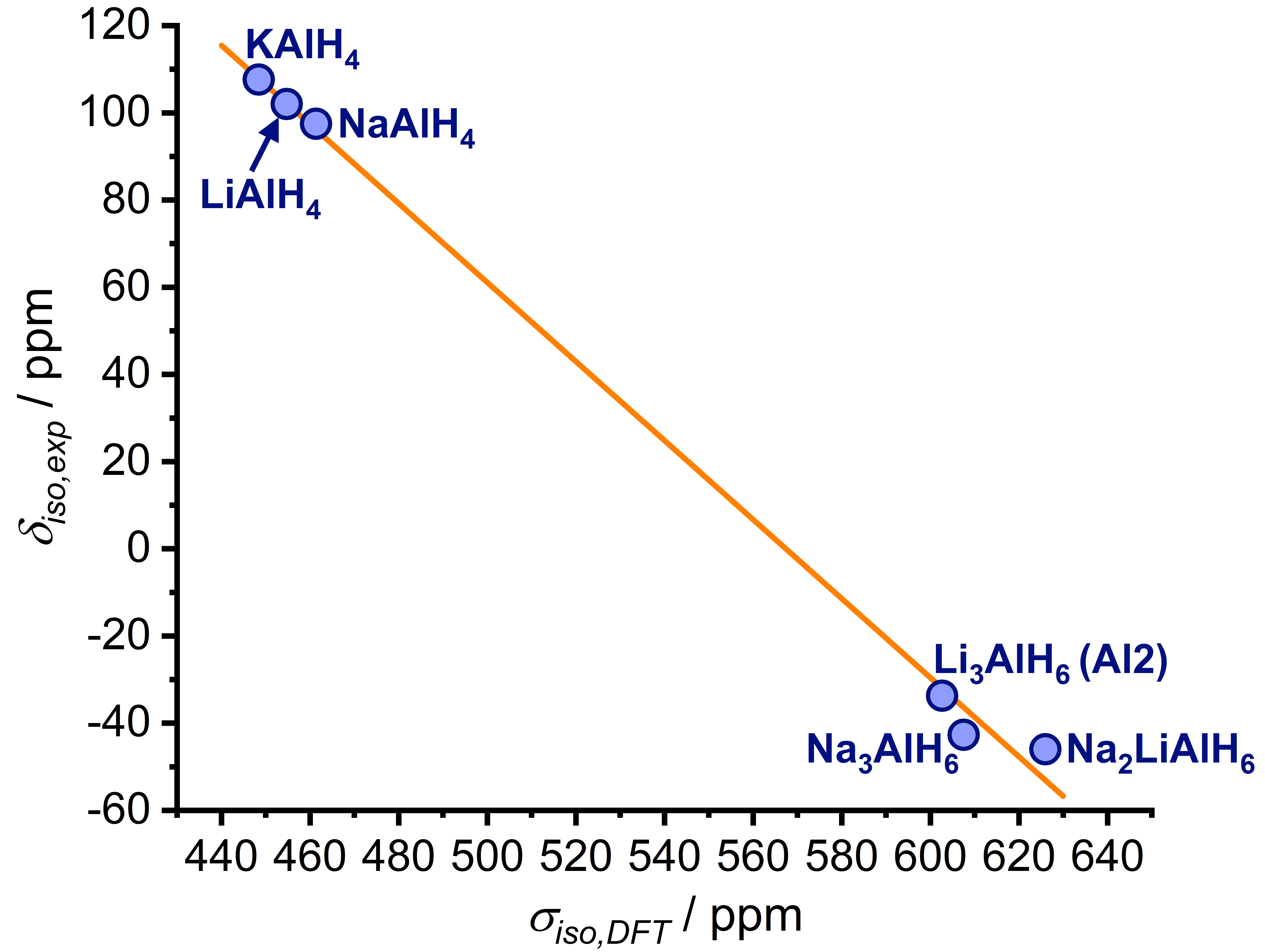}
  \caption{{Experimentally determined $^{27}$Al chemical shifts $\delta_{iso}$ vs isotropic shieldings $\sigma_{iso}$ obtained from the DFT-GIPAW computations for some alkali aluminium hydrides. The squared correlation coefficient $R^2$ of the fit is 0.997.}}
  \label{fgr:delta_sigma}
\end{figure}

To establish an equation to convert the shielding $\sigma_{iso,DFT}$ into chemical shifts $\delta_{iso,DFT}$, a linear regression was computed across all six systems. Literature data for  $\delta_{iso}$ of the two hydrides studied here, \ce{NaAlH4} and \ce{KAlH4}, were included in the regression.\cite{Zhang09,ZF19} The data points as well as the linear fit are shown in Fig.~\ref{fgr:delta_sigma}.  The resulting equation for the conversion is: 
\begin{equation}
  \delta_{iso,DFT,1} = 514.3\,\text{ppm} - 0.9064 \cdot \sigma_{iso,DFT}\text{.}
   \label{eq:delta2}
\end{equation}

\section{Results and Discussion}
\subsection{Experimental results for \ce{KAlH4}}

\begin{figure}
  \includegraphics[width=8.3cm]{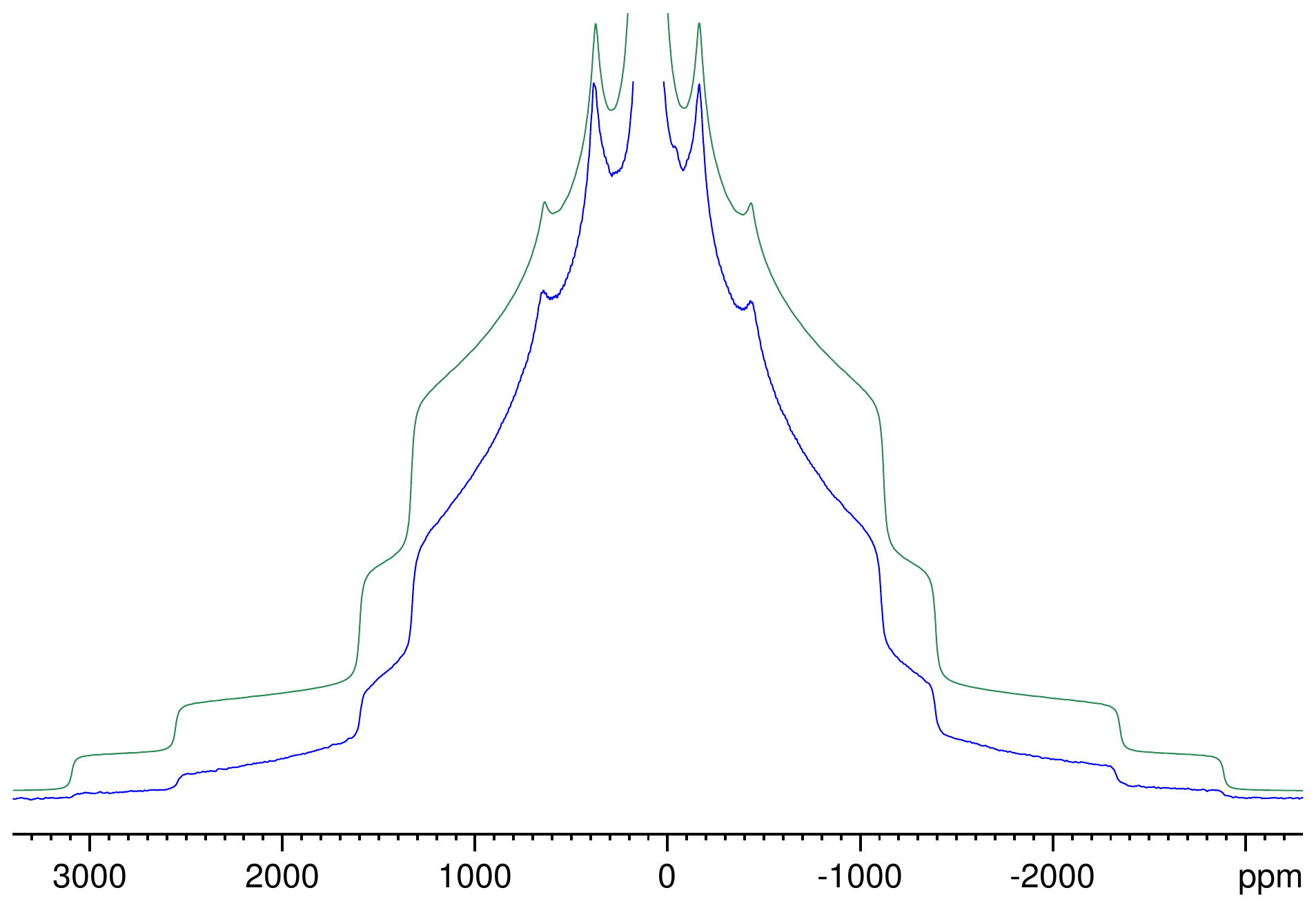}
  \caption{Experimental $^{27}$Al NMR spectrum  of \ce{KAlH4} obtained from a Solomon echo generated with a pulse spacing of 600\,$\mu$s (blue) and the best fit of this spectrum (green) using the following parameters: $\delta_{iso}=108$\,ppm, $C_Q = 1.30$\,MHz and $\eta = 0.64$. For the experimental spectrum, a Lorentzian line broadening with $LB= 1200$\,Hz was applied to suppress small modulations caused by the echo at 2$\tau$ that was still noticeable despite the long pulse spacing used. The spectrum is cut off at about 6\% of the maximum intensity of the central line. For the simulated spectrum, a Lorentzian line broadening with $LB= 2000$\,Hz was applied. }
  \label{fgr:KAlH4}
\end{figure}

Fig.~\ref{fgr:KAlH4} shows a $^{27}$Al~NMR spectrum obtained as a Fourier transform of the Solomon echo (formed at $\tau$ after the second pulse) for a stationary \ce{KAlH4} sample in comparison with the best fit. The depicted lineshape is typical of a first-order quadrupolar broadened  powder spectrum. \cite{Man11} The positions of the characteristic discontinuities of the lineshape are obviously nicely reproduced by the fit. The significantly reduced intensity in the outer wings of the experimental spectrum is mainly caused by the insufficient excitation width of the finite pulses used. \cite{Freude00} Neither the finite excitation width nor the finite bandwidth of the NMR probe are taken into account by the simulation program used. It should be noted that the resonance line of the central transition is not caused by the Solomon echo but by the free induction decay following the second pulse of the pulse sequence.\cite{Man00}

The fact that such a complex resonance line can be described by only three parameters is somewhat amazing: the position of the line is mainly determined by the isotropic chemical shift $\delta_{iso}$ and its shape is governed by the quadrupole coupling constant $C_Q$ and the asymmetry parameter $\eta$. \cite{Man11} The latter two parameters also have a small but significant effect on the line position via the quadrupole induced shift $\delta_{qis}$.\cite{Samoson85} This second-order effect that is inversely  proportional to the square of the magnetic field applied influences the centre of gravity of the central transition and each pair of satellite transitions in a characteristic way. Hence, the line position read off a spectrum must generally be corrected to obtain the true chemical shift for any half-integer quadrupolar nucleus. Simulation programs take the quadrupole induced shift into account.

It should be noted that from the NMR measurements reported here, only the magnitude, but not the sign of the quadrupole coupling constant $C_Q$ can be determined. The resonance lines of every pair of satellite transitions are mirror images of each other. In accordance with most of the NMR literature, \cite{quadnmr} we omit the absolute value bars for $C_Q$ throughout the paper.

To have a suitable simulation program at hand is useful, but not essential for determining the spectral parameters from experimental spectra measured for stationary samples. In fact, rather good estimates of these parameters can directly be read off the experimental spectrum in Fig.~\ref{fgr:KAlH4}. The maxima of the inner satellite transitions at 380 and $-162$\,ppm yield a value of 109\,ppm for the centre of gravity of these transitions.
Since for any nucleus with $I = \tfrac {5}{2}$, the position of the  centre of gravity of the inner satellite transitions is known to be much closer to the isotropic chemical shift than the centre of gravity of the central transition,\cite{Samoson85} this estimate is very close to the true isotropic chemical shift $\delta_{iso}=(108\pm2)$\,ppm found by means of the simulation program.

	The quadrupole coupling constant $C_Q$ can be derived from the total spread of the satellite transitions $\Delta\nu_{TS}(m)$. For the inner satellite transitions ($m=\tfrac {3}{2}$), the following relation holds: \cite{Taylor75,Freude00,ZF19}  
\begin{equation}
  C_Q=\frac{10}{3}\Delta\nu_{TS}(\tfrac {3}{2}).
   \label{eq:a}
\end{equation}
Since the outermost shoulders of the inner satellite transitions are localized at 1610\,ppm (low-field) and $-1390$\,ppm (high-field), we have a total spread of 3000\,ppm corresponding to $\Delta\nu_{TS}(\tfrac {3}{2})\approx390$\,kHz. With eqn~(\ref{eq:a}), this value leads to the rather good estimate of $C_Q\approx1.3$\,MHz.  The asymmetry parameter can be estimated from the ratio of the splitting of the maxima $\Delta\nu_M(m)$ to the total spread of any pair of satellite transitions according to the following equation:\cite{ZF19}
\begin{equation}
  \eta=1-\frac{2\Delta\nu_M(m)}{\Delta\nu_{TS}(m)}.
   \label{eq:b}
\end{equation}
With the above given values for the positions of the maxima of the inner satellite transitions, we find a splitting of 542\,ppm corresponding to $\Delta\nu_M(\tfrac {3}{2})\approx71$\,kHz. With eqn~(\ref{eq:b}), we obtain $\eta\approx0.64$. Of course, the quadrupole parameters can also be obtained from the outer satellite transitions, but the discontinuities are usually much better defined for the narrower inner satellite transitions.
The estimates for $C_Q$ and $\eta$ turned out to be so precise that they could not be improved by means of the simulation program. We think that the margins of error given for $C_Q$ ($\pm20$\,kHz) and $\eta$ ($\pm0.02$) in Table \ref{tbl:EXP_vs_GIPAW}  are rather conservative estimates. Hence, the reliability of determining the parameters of the quadrupole coupling for \ce{KAlH4} from static spectra is as good as that from spectra measured under MAS conditions. \cite{ZF19} Regarding the accuracy of the determination of the isotropic chemical shift, the MAS approach is definitely superior. 

\begin{figure}
	\includegraphics[width=8.3cm]{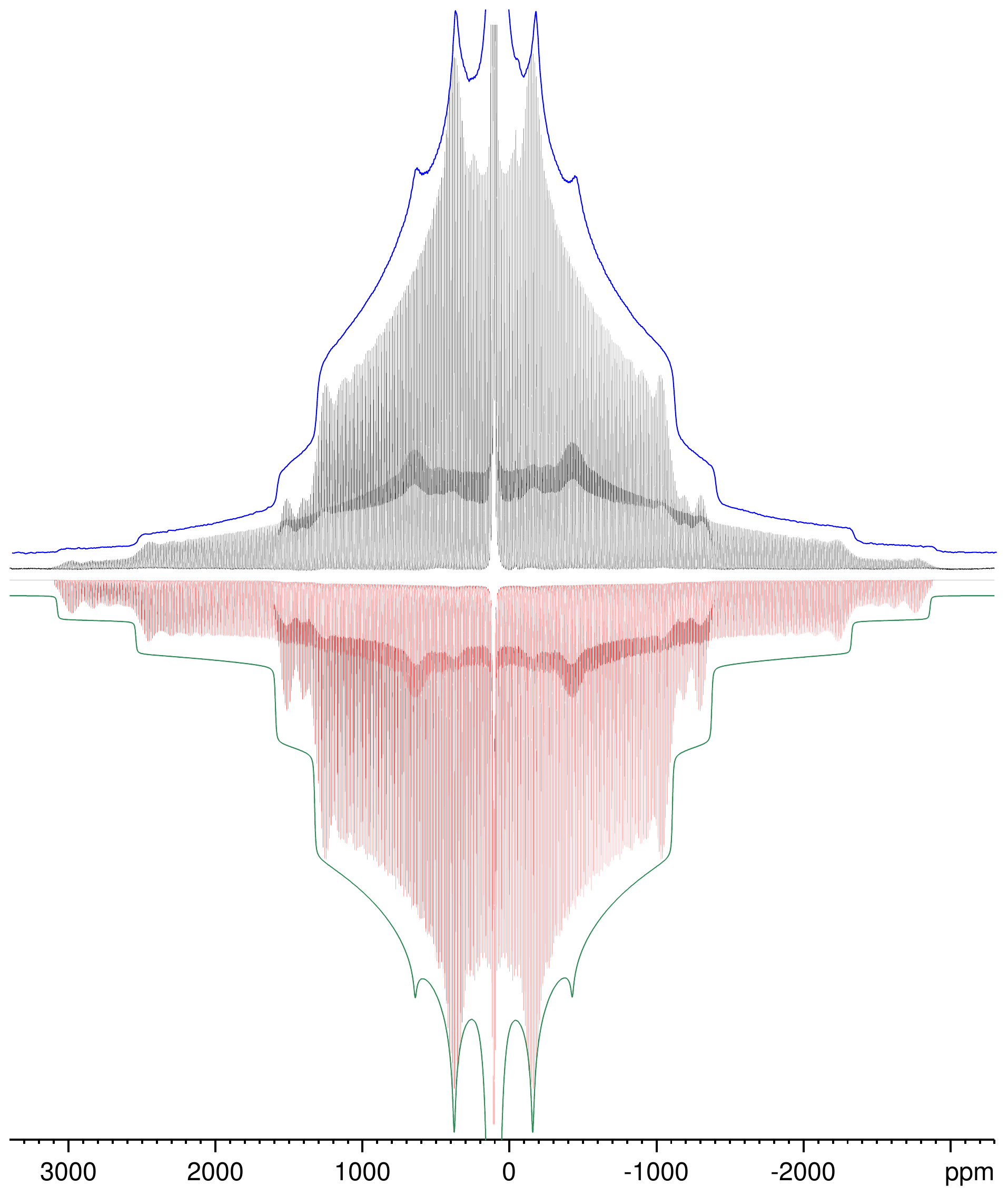}
  \caption{Experimental $^{27}$Al~NMR spectra of  measured for a stationary sample (blue) and the same sample spinning at  $\nu_{MAS} = 1.4$\,kHz (black) and their simulations (green and red, respectively) with the parameters $\delta_{iso}=107.6$\,ppm, $C_Q = 1.29$\,MHz and $\eta = 0.64$. To ease the comparison, the simulated spectra have been inverted.}
  \label{fgr:KAlH4_T}
\end{figure}

Fig.~\ref{fgr:KAlH4_T} compares the $^{27}$Al~NMR spectra measured with and without MAS with the simulated spectra using the parameters previously determined from MAS spectra measured at different spinning speeds. The quality of the fits can better be judged from the enlarged version of Fig.~\ref{fgr:KAlH4_T} given as Fig.\,S2 in the Supporting Information. Amazingly, the tiny amount of the cryolite-like by-product \ce{Na3AlH6} detected as narrow line at $-42.7$\,ppm in the MAS NMR spectrum \cite{ZF19} can also be identified  as shoulder at about $-40$\,ppm in the static spectrum. 

Although the lineshape simulations neglect the effect of a possible anisotropy of the  chemical shift, they reproduce the experimental lineshapes quite well. Hence, any anisotropy must be rather small and contributes only to the line broadening of the static spectrum. This observation is in line with the results of our GIPAW calculations.  For the span $\Omega$, a parameter describing the total spread of a resonance line governed by chemical shift anisotropy, \citep{Mason93} these calculations yield a value of 11\,ppm corresponding to less than 1.5\,kHz (see Supporting Information, Table S1). 

\subsection{Experimental results for \ce{NaAlH4}}

\begin{figure}
  \includegraphics[width=8.3cm]{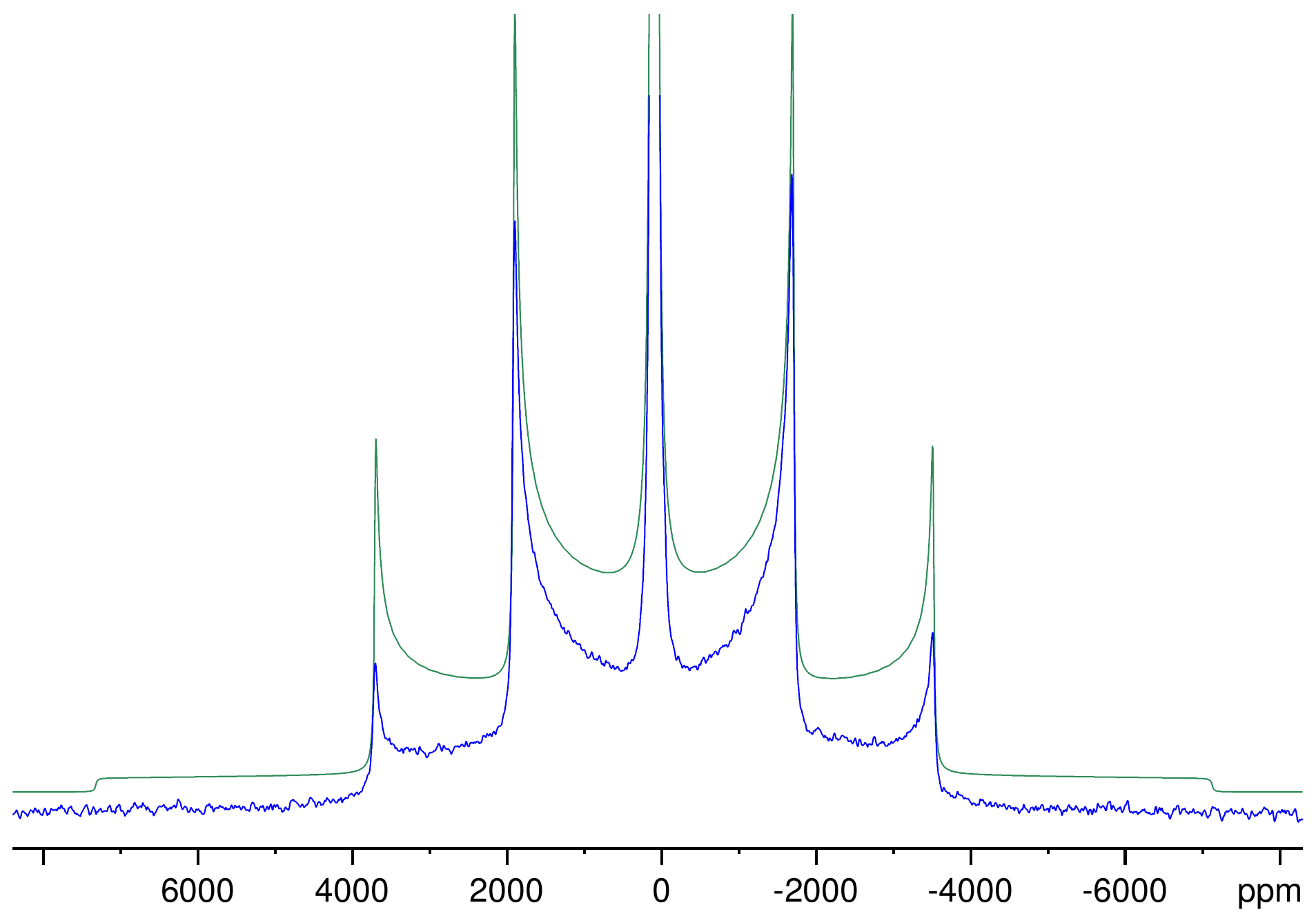}
  \caption{Experimental $^{27}$Al NMR spectrum  of \ce{NaAlH4} obtained from a Solomon echo generated with a pulse spacing of 600\,$\mu$s (blue) and the best fit of this spectrum (green) using the following parameters: $\delta_{iso}=99$\,ppm, $C_Q = 3.11$\,MHz and $\eta = 0$. Lorentzian line broadening with $LB= 3000$\,Hz and $LB= 4000$\,Hz was applied for the experimental spectrum and the simulated spectrum, respectively. The experimental spectrum is cut off at about 12\% of the maximum intensity of the central line.}
  \label{fgr:NaAlH4}
   \end{figure}
  
Fig.~\ref{fgr:NaAlH4} shows a $^{27}$Al~NMR spectrum obtained as a Fourier transform of the Solomon echo for a stationary \ce{NaAlH4} sample in comparison with the best fit. Again, the lineshape is typical of a first-order quadrupolar broadened powder spectrum, but this time for the case of an axially symmetric electric field gradient tensor, i.e.\ $\eta=0$. \cite{FreudeHaase, Freude00, Man11} Apart from the central transition, the spectrum shows the sharp maxima of the four satellite transitions. 
These maxima correspond to the perpendicular components of the axially symmetric efg tensor. The maxima of the inner satellite transitions are found at about 1887 and $-1674$\,ppm,  corresponding to a splitting of the maxima $\Delta\nu_M(\tfrac {3}{2})$ of about 464\,kHz. Since, for $\eta=0$, the total spread  $\Delta\nu_{TS}(m)$ equals $2\Delta\nu_{M}(m)$ (see eqn~(\ref{eq:b})), we obtain with eqn~(\ref{eq:a}) a value of $C_Q\approx3.09$\,MHz as an estimate of the strength of the quadrupole coupling. Using the simulation program, we found for the best fit $C_Q=3.11$\,MHz. 

As expected from the known structure of \ce{NaAlH4},\cite{Ozolins2004, Hauback03} we found no indication for a deviation from axial symmetry. Since the Al atoms are located in Wyckoff position 4b, i.e.\ on a $C_4$ axis, the asymmetry parameter $\eta$ is bound to be zero. However, with the finite linewidth observed, it is difficult to verify experimentally that $\eta=0$ holds. From our experimental data, we can safely deduce that $\eta<0.01$ is the upper bound for a deviation from axial symmetry.

Because of the enormous width of the spectrum, the deviations of the experimental from the simulated lineshape are more pronounced than in the case of \ce{KAlH4}. From the above given splitting of the maxima of the inner transition $\Delta\nu_M(\tfrac {3}{2})$, it follows that the total spread of the outer satellite transition $\Delta\nu_{TS}(\tfrac {5}{2})$ is about 1.86\,MHz. In particular, the shoulders of the outer satellite transition can not be discerned in the experimental spectrum. However, these distortions of the lineshape are less severe than for spectra obtained from the free induction decay after a single pulse. During the inevitable receiver dead time after the pulse, the fast decaying signal corresponding to the broad features is lost and only the sharp maxima can be seen in these spectra. \cite{Tarasov97, Verkuijlen09} It seems unlikely that this approach could produce meaningful results for the quadrupole parameters for the general case $\eta\neq0$. In our own attempts to record $^{27}$Al~NMR spectra for   stationary \ce{NaAlH4} samples by single pulse excitation, we ran into serious phasing problems. 

\begin{figure}
  \includegraphics[width=8.3cm]{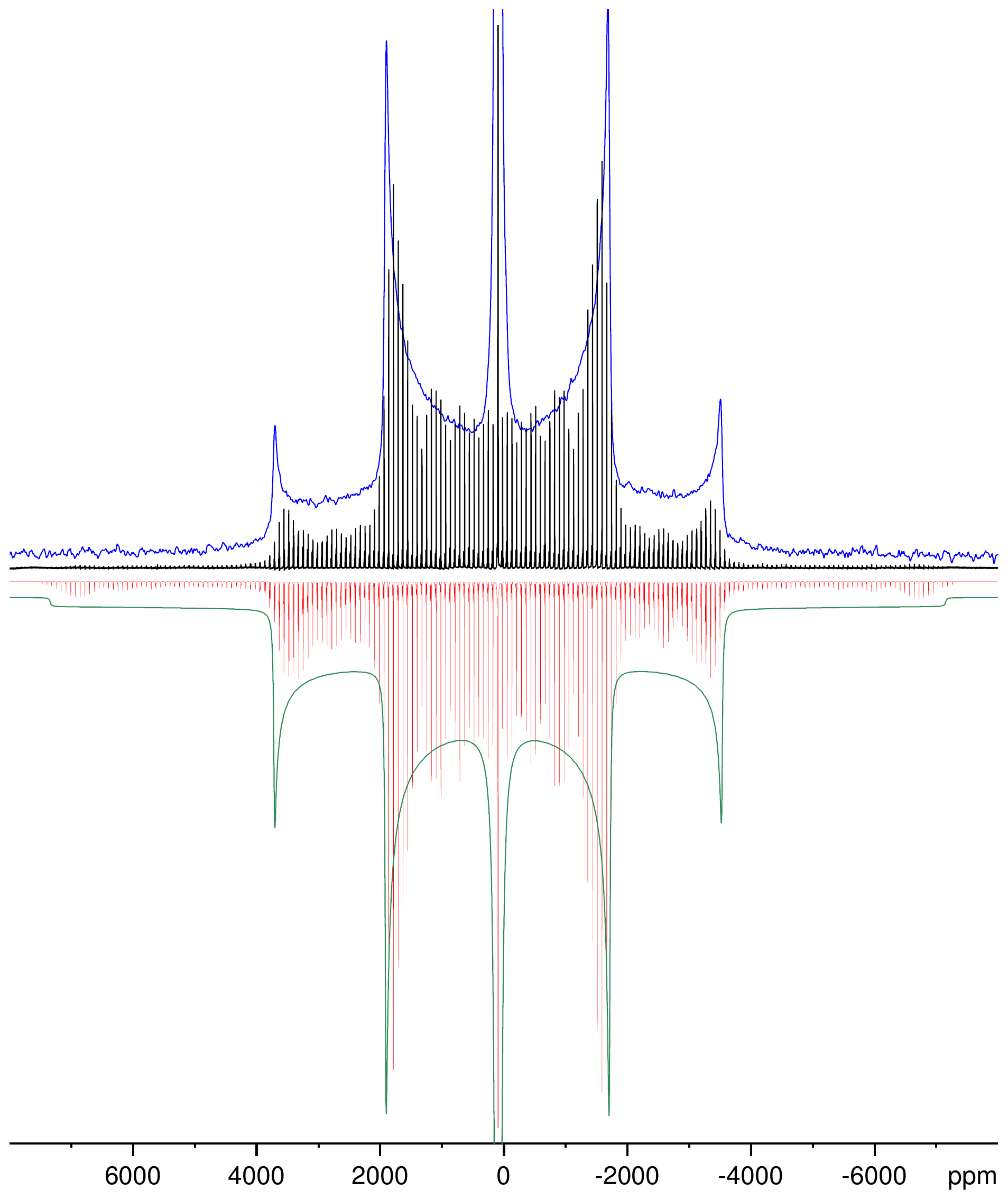}
  \caption{Experimental $^{27}$Al~NMR spectra of \ce{NaAlH4} measured for a stationary sample (blue) and the same sample spinning at  $\nu_{MAS} = 10$\,kHz (black) and their simulations (green and red, respectively) with the parameters $\delta_{iso}=97.2$\,ppm, $C_Q = 3.11$\,MHz and $\eta = 0$. To ease the comparison, the simulated spectra have been inverted.}
  \label{fgr:NaAlH4_T}
\end{figure}

Fig.~\ref{fgr:NaAlH4_T} compares the $^{27}$Al~NMR spectra measured with and without MAS with the simulated spectra using the parameters determined from MAS spectra measured at different spinning speeds. The quality of the fits can better be judged from the enlarged version of Fig.~\ref{fgr:NaAlH4_T} given as Fig.\,S3 in the Supporting Information. As for \ce{KAlH4}, we do not see any of the characteristic effects of a possible anisotropy of the chemical shift in the spectra recorded for stationary samples or under MAS conditions. Our GIPAW calculations yield a value of about 19\,ppm for the span $\Omega$, corresponding to about 2.5\,kHz.

\begin{figure}
  \includegraphics[width=8.3cm]{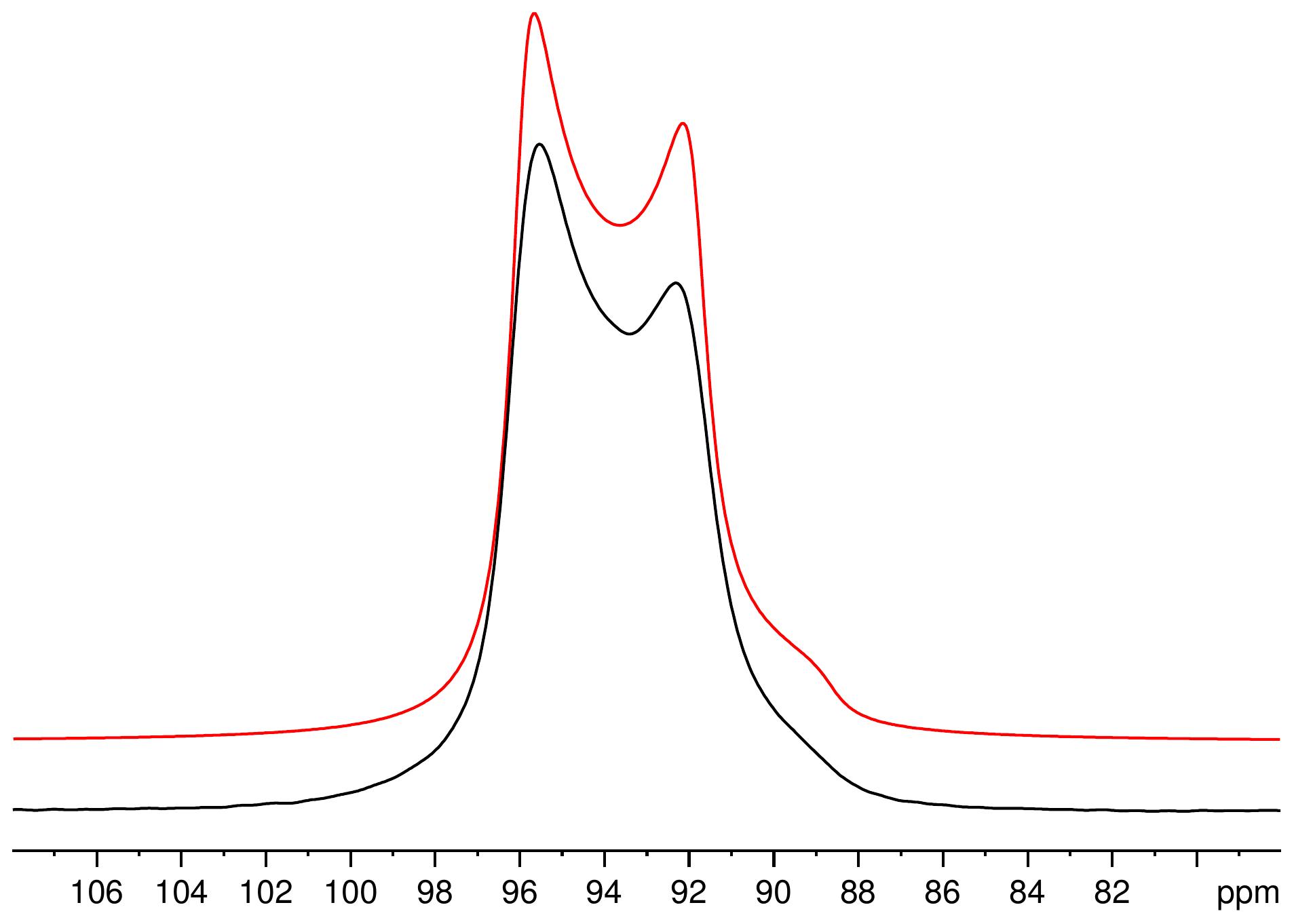}
  \caption{Experimental $^{27}$Al~NMR spectrum of \ce{NaAlH4} measured  at  $\nu_{MAS} = 5$\,kHz (black) and its simulation (red) with the parameters $\delta_{iso}=97.2$\,ppm, $C_Q = 3.11$\,MHz and $\eta = 0$. Only the region of the centreband is shown. Lorentzian line broadening with $LB= 20$\,Hz and $LB= 120$\,Hz was applied for the experimental spectrum and the simulated spectrum, respectively.}
  \label{fgr:NaAlH4_CS}
\end{figure}

As distinct from the case of \ce{KAlH4} at the same field strength,\cite{ZF19} the effect of second-order quadrupole interaction is not fully masked by other broadening mechanisms. The centreband of the central transition shown in Fig.~\ref{fgr:NaAlH4_CS} exhibits two maxima separated by about 3\,ppm (95.6 and 92.3 ppm). These maxima  are best resolved at spinning rates between 3 and 6 kHz. We assume that the reduced resolution at higher spinning rates is caused by frictional heating that leads to a significant increase of the temperature in certain parts of the sample that influences the chemical shift and/or the quadrupole coupling parameters.\cite{Brus00, Antonijevic05} 
Fig.~\ref{fgr:NaAlH4_CS} also shows that the splitting and the general lineshape of the centreband can nicely be fitted using the same parameters as for the simulated spectra of the satellite transitions in Fig.~\ref{fgr:NaAlH4_T}. However, at the still rather low ratio of the quadrupole coupling constant to the resonance frequency ($C_Q/\nu_L\approx0.024$), the experimental lineshape of the centreband of the central transition does not really show the discontinuities characteristic for second-order quadrupolar broadening.\cite{FreudeHaase, Freude00, Man11} A reliable determination of the parameters of the quadrupole coupling would not be possible. These parameters can be obtained with much higher accuracy from the satellite transitions governed by first-order quadrupole interaction, either from spectra of stationary samples measured by means of the Solomon echo sequence or from the characteristic sideband patterns of MAS spectra measured with single-pulse excitation. Since the second-order broadening is proportional to the  square of the quadrupole coupling constant, it can, of course, not deliver any information about the sign of $C_Q$ either.

\subsection{GIPAW vs experiment}

The results from the DFT-GIPAW calculations are collected in Table \ref{tbl:EXP_vs_GIPAW}, where experimental values from this study and previous work are given for comparison. With regard to the  $^{27}$Al chemical shifts, both the use of \ce{LiAlH4} as sole reference system (eqn~(\ref{eq:delta1})) and the calculation by means of a linear regression over several alkali aluminium hydrides  (eqn~(\ref{eq:delta2})) deliver very good agreement with experiment. While the former approach should be suitable for converting DFT-computed chemical shieldings into chemical shifts for structurally closely related systems, the regression-based approach can be expected to be more broadly applicable for diverse alkali aluminium hydrides. The direct results of the DFT-GIPAW calculations ($C_Q$, $\eta$, $\sigma_{ii}$) for all substances mentioned are summarised in Table S1 (Supporting Information).

The DFT-GIPAW quadrupole coupling constant $C_Q$ of \ce{KAlH4} is approximately 30\% (0.4 MHz) larger than the experimental values. A similar overestimation was found in a previous DFT-GIPAW study of \ce{Na3AlH6}.\cite{Zhang09} In that study, the observed difference was attributed to the influence of thermal motion on the electric field gradients, which is not taken into account in static DFT calculations. In contrast, our calculated quadrupole coupling constant for \ce{NaAlH4} is in perfect agreement with the experimental one. It is important to note that our procedure is different from that applied by Zhang et al. \cite{Zhang09}. These authors used the experimentally determined value for $C_Q$ and the calculated field gradient of \ce{NaAlH4} to determine the quadrupole moment $Q$($^{27}$Al). The thus obtained quadrupole moment was then used to calculate the quadrupole couplings for the other alanates investigated. Our calculations of the parameters of the quadrupole interaction are solely based on the crystallographic data and the generally accepted value of the quadrupole moment $Q$($^{27}$Al)$= 146.6$\,mbarn.\cite{Pyykko18} Hence, the calculated data reported here are fully independent from our experimental data. 

Recently, a slightly higher value of the quadrupole moment $Q$($^{27}$Al)$= 148.2$\,mbarn was recommended on the basis of highly accurate coupled cluster calculations with single, double, and perturbative triple excitations [CCSD(T)] for Al-containing molecules.
 \cite{Aerts19} By using this value, the magnitudes of the calculated quadrupole coupling constants would be about 1\% higher than given in Table \ref{tbl:EXP_vs_GIPAW} and Table S1 (Supporting Information). DFT calculations carried out with the newly recommended value for the quadrupole moment typically resulted in $C_Q$ values deviating by about 10 to 15\% from experiment.\cite{Aerts19}

With regard to the asymmetry parameters, the DFT-calculated $\eta = 0.59$ for \ce{KAlH4} agrees very well with the experimental values, both from the literature and from the present work. For \ce{NaAlH4}, $\eta = 0$ reflects the high symmetry of the local environment (site symmetry: $\bar{4}$).\cite{Hauback03, Ozolins2004} All four Al--H bonds are equivalent by symmetry.

Although not in the focus of the present paper, we want to mention the GIPAW results for the counterions. We obtained a good agreement between experiment and calculations for the quadrupole parameters for $^{39}$K in \ce{KAlH4}: $C_Q=0.562/0.68$\,MHz (Exp.\cite{ZF19}/GIPAW) and  $\eta=0.74/0.79$. Unfortunately, the result for $^{23}$Na in \ce{NaAlH4} is less encouraging: $C_Q=0.15$/$-0.46$\,MHz (Exp.\cite{Zhang09}/GIPAW) and  $\eta=0/0.$ As mentioned above, we omit the sign of $C_Q$ for the experimental data.

\begin{table}
  \caption{Comparison of the results of DFT-GIPAW calculations  with the experimental data from NMR spectra measured with and without MAS using either single-pulse excitation (SPE) or the Solomon echo pulse sequence (SE). For $\delta_{iso}$ from GIPAW calculations, both $\delta_{iso,DFT,1}$ (eqn~(\ref{eq:delta1})) and $\delta_{iso,DFT,2}$ (eqn~(\ref{eq:delta2})) are given.}
  \label{tbl:EXP_vs_GIPAW}
  \begin{tabular}{llccc}
    \hline
    Material  & Method &  $\delta_{iso}$/ppm & $C_Q$/MHz & $\eta$\\
    \hline
    \ce{KAlH4}  & SPE, MAS\cite{ZF19} & $107.6\pm0.2$ & $1.29\pm0.02$ &  $0.64\pm0.02$  \\
     &  SE, static & $108\pm2$ & $1.30\pm0.02$ & $0.64\pm0.02$\\
     &  GIPAW & 108.4/107.8 & 1.69 & 0.59\\
    \ce{NaAlH4}  & SPE, MAS\cite{Zhang09} & 97.5 & 3.15 & 0.04\\
     & SPE, MAS & $97.2\pm0.3$ & $3.11\pm0.03$ & $<$0.05 \\
     & SPE, static\cite{Tarasov97} & $101\pm3$ & 3.08 & 0\\
     & SPE, static\cite{Verkuijlen09} & - & $3.10\pm0.05$ & $0.00\pm0.05$\\
     & SE, static & $99\pm3$ & $3.11\pm0.02$ & $<$0.01\\
     & GIPAW & 95.4/96.1 & 3.14 & 0\\
    \hline
  \end{tabular}
\end{table}

\section{Conclusions}

For two complex aluminium hydrides, we have shown that by using the Solomon echo sequence the parameters of the quadrupole coupling can be determined from $^{27}$Al NMR spectra measured for stationary samples with at least the same precision as from MAS NMR spectra. For the isotropic chemical shift, the precision of the MAS approach is superior.

The avoidance of mechanical stress and frictional heating caused by fast sample spinning is certainly a great advantage of NMR measurements of stationary samples, in particular when less stable materials are to be investigated. 

Furthermore, the Solomon echo sequence can be used with any NMR probe that can generate sufficiently strong rf pulses. Obviously, \textit{in situ} studies of phase transitions or reactions at high pressure in a wide temperature range are easier to accomplish for stationary samples than for those under MAS conditions.

The use of Solomon echoes is of course not limited to $^{27}$Al in aluminium hydrides, but should be feasible for other half-integer quadrupolar nuclei with small or moderate quadrupole couplings in materials from various classes of substances, provided that the longitudinal relaxation is slow enough for intense Solomon echoes to be formed.

The encouraging agreement between experimental and GIPAW data obtained is in line with the general notion that modern density functional theory is accurate enough to provide a good description of the electronic structure and hence the efg and quadrupole coupling in a very wide range of solids.\cite{Zwanziger12}

The sensitivity of the quadrupole interaction to the local geometry offers a valuable source of information about the structure. Hence, by determining the parameters of the quadrupole coupling, both from NMR measurements and DFT calculations, it should be possible to distinguish between different structural models for new materials based on diffraction data.

\subsection{Conflict of interest}
There are no conflicts to declare.

\begin{acknowledgement}

The authors thank Dr Michael Felderhoff, MPI f\"ur Kohlenforschung in M\"ulheim an der Ruhr, for the preparation of the aluminium hydride samples used. MF gratefully acknowledges funding by the Deutsche Forschungsgemeinschaft (German Research Foundation, DFG) through a Heisenberg fellowship (project no. 455871835). 

\end{acknowledgement}

\subsection{Supporting Information}
Supporting Information is available. 

\subsection{ORCID}
Bodo Zibrowius: 0000-0002-3894-7318\\
Michael Fischer: 0000-0001-5133-1537

\bibliography{Solomon}

\end{document}